\documentstyle[axodraw]{elsart}
\newcommand{\ee}{\mbox{$e^-e^-$}}
\newcommand{\ep}{\mbox{$e^+e^-$}}
\newcommand{\eg}{\mbox{$e^-\gamma$}}
\newcommand{\pp}{\mbox{$\gamma\gamma$}}
\newcommand{\eezz}{\mbox{$e^-e^-Z^0Z^0$}}
\newcommand{\enzw}{\mbox{$e^-\nu_eZ^0W^-$}}
\newcommand{\nnww}{\mbox{$\nu_e\nu_eW^-W^-$}}
\newcommand{\ie}{{\em i.e.}}
\newcommand{\xs}{cross section}
\newcommand{\bkg}{background}
\newcommand{\sm}{standard model}
\newcommand{\br}{branching ratio}
\newcommand{\EW}{electroweak}
\newcommand{\EM}{electromagnetic}
\newcommand{\lc}{linear collider}
\newcommand{\cm}{centre of mass}
\newcommand{\aqgc}{anomalous quartic gauge couplings}
\newcommand{\beq}{\begin{equation}}
\newcommand{\eeq}{\end{equation}}
\newcommand{\beqa}{\begin{eqnarray}}
\newcommand{\eeqa}{\end{eqnarray}}

\begin{document}

\begin{flushright}
hep-ph/9411277\\
MPI-PhT/94-78\\
LMU-20/94\\
November 1994
\end{flushright}
\vskip2cm

\begin{frontmatter}
\title{Probing Anomalous Quartic Vector Boson Couplings
       in $e^-e^-$ Collisions}
\author{Frank Cuypers}
\address{{\tt CUYPERS@MPPMU.MPG.DE}\\
	Max-Planck-Institut f\"ur Physik,
	Werner-Heisenberg-Institut,
	D--80805 M\"unchen, Germany}
\author{Karol Ko\l odziej\thanksref{work}\thanksref{leave}}
\address{{\tt KAROL@HEP.PHYSIK.UNI-MUENCHEN.DE}\\
	Theoretische Physik,
	Ludwig-Maximilians-Universit\"at,
	Theresienstra\ss e 37,
	D--80333 M\"unchen, Germany}
\thanks[work]{Work supported by the German Federal Ministry for Research
        and Technology under contract No.~05 6MU 93P}
\thanks[leave]{On leave from the Institute of Physics, University of Silesia,
        ul.~Uniwersytecka 4, PL--40007 Katowice, Poland}

\begin{abstract}
We study the incidence of genuine quartic vector boson couplings
in the reactions $\ee\to\eezz$, \enzw\ and \nnww.
We find that a linear collider operated in the \ee\ mode
will provide bounds on these couplings
which are competitive with those
to be obtained in the \ep\ mode.
\end{abstract}
\end{frontmatter}
\clearpage

Quadrilinear gauge boson couplings
are an essential ingredient of the \sm\ of \EW\ interactions.
They occur naturally in any Yang-Mills theory,
where their form is determined by the gauge symmetry of the theory.
On the other hand,
in an effective field theory scenario,
quadrilinear vector boson coupling are usually introduced
to parametrize {\em `new physics'} effects
such as the exchange of some heavy particles
or loop effects of some unified gauge theory.
They can also be introduced
in order to guaranty the tree level unitarity
of such an effective theory \cite{fred}.
In contrast to trilinear vector boson couplings,
which will already be tested directly at LEP II,
the quadrilinear ones will remain unexplored
until the next generation of accelerators becomes operational.

Even existing data is already sensitive to Yang-Mills couplings
through loop effects \cite{Schildknecht}.
The \sm\ tests performed by the LEP I experiments
leave no much room for dramatic deviations from the \sm.
 From this point of view it seems reasonable
to restrict oneself to those deviations
from the Yang-Mills form of the couplings
which leave intact as much of the \sm\ structure as possible.
Therefore, in the following,
we will only consider anomalous quartic couplings
between massive vector bosons
and assume the trilinear vector boson couplings
to take on their \sm\ form.

There are infinitely many operators
which can induce non-standard quadrilinear vector boson couplings.
However,
we concentrate here on those which have the lowest dimension
and which do not induce non-standard trilinear couplings
at the same time.
There are only two such ``genuine'' quartic operators
of dimension four.
They add the following pieces to the \sm\ lagrangian \cite{bb}:
\newcommand{\cwt}{\cos^2\theta_w}
\newcommand{\cwf}{\cos^4\theta_w}
\beqa
{\cal L}_0 = g_0 g_W^2
& \Bigl( & 		W^{+\mu} W^-_\mu W^{+\nu} W^-_\nu \nonumber\\
& + &	 {1\over\cwt}	W^{+\mu} W^-_\mu Z^\nu Z_\nu
+	 {1\over4\cwf}	Z^\mu Z_\mu Z^\nu Z_\nu
\Bigr)\ ,
\label{e1}
\\
{\cal L}_c = g_c g_W^2
& \Bigl( & {1\over2}	(W^{+\mu} W^-_\mu W^{+\nu} W^-_\nu
+			W^{+\mu} W^+_\mu W^{-\nu} W^-_\nu) \nonumber\\
& + &	 {1\over\cwt}	Z^\mu W^+_\mu Z^\nu W^-_\nu
+	 {1\over4\cwf}	Z^\mu Z_\mu Z^\nu Z_\nu
\Bigr)\ .
\label{e2}
\eeqa
The corresponding Feynman rules
for the two modified $WWWW$ and $WWZZ$ vertices
and the new $ZZZZ$ vertex
are shown in Fig.~\ref{f1}.
Note that anomalous quartic operators involving a photon
must have at least dimension six,
since they must contain derivatives of the photon field
in order to conserve \EM\ gauge invariance.
We do not consider such operators here.

These \aqgc\
have been shown to be testable in \ep\ collisions
at a \lc\ of the next generation \cite{bb}.
Several world wide collaborations
(CLIC, DESY, JLC, NLC, TESLA, VLEPP)
are currently developing designs
for machines capable of colliding electron and positron beams
at a \cm\ energy of 500 GeV
with an integrated luminosity of 10 fb.
Colliding electrons with electrons
is an important complementary experiment
which can be performed at such a \lc.
Indeed,
the currently established discovery potential of \ee\ scattering
is appreciable \cite{zeuthen}.
Essentially,
this is due to the low level of \sm\ \bkg\
and to the possibility of polarizing both beams.
In particular,
\ee\ collisions have been shown to be as sensitive a probe
of anomalous trilinear gauge couplings
as \ep, \eg\ and \pp\ collisions \cite{cc}.
In conjunction with these other experiments,
the \ee\ running mode provides
important additional discriminating power
if several anomalous couplings take non-vanishing values at the same time.
As we shall show in the next paragraphs,
the same is true for
the genuine \aqgc.

We have computed the effect of the new interactions represented by extra
terms (\ref{e1}, \ref{e2}) in the \sm\ lagrangian
on the total \xs s of the reactions
\beqa
\ee & \to & \nnww \label{e3} \ , \\
\ee & \to & \enzw \label{e4} \ , \\
\ee & \to & \eezz \label{e5} \ .
\eeqa
The process (\ref{e5}) takes place
for all three combination of polarization of the initial electron beams,
$LL$, $LR$ and $RR$.
In contrast,
the reaction (\ref{e4})
occurs only in the $LL$ and $LR$ modes
while the reaction (\ref{e3}) can take place solely with $LL$ polarization.
We set the electron mass equal to zero throughout.

The \sm\ rates for reactions (\ref{e3}--\ref{e5})
are rather low but observable
because there is no hadronic \bkg\
as in \ep\ collisions \cite{ck}.
As it turns out,
the low values of the \sm\ \xs s
are due to very powerful cancellations
between different diagrams.
Considering the large number of diagrams
contributing to each reaction
%(in the Feynman gauge,
%86 for reaction (\ref{e5}),
%88 for reaction (\ref{e4}) and
%66 for reaction (\ref{e3}))
this is a highly non-trivial phenomenon
which is to be traced back to the miraculous properties
of a renormalizable theory.
The quartic gauge coupling
play in this respect an essential role
in reactions (\ref{e3}, \ref{e4}).

However, as soon
as anomalous couplings are introduced,
the cancellations between diagrams are not as effective
and unitarity is spoiled.
Even for very small values of $g_0$ or $g_c$
the \xs s for processes (\ref{e3}--\ref{e5})
increase significantly,
the more so at higher energies.
In processes (\ref{e3},\ref{e4})
the genuine \aqgc s modify the diagrams
involving the first and second vertices of Fig.~\ref{f1}, respectively.
In reaction (\ref{e5})
they participate
via a new diagram involving the third vertex of Fig.~\ref{f1}.

Our results are shown in Figs~\ref{f2}\ and \ref{f3} where
we have plotted
the 95\%\ confidence level
contours of the function
\beq
\chi^2 = \left( {n(g_0,g_c)-n_{SM} \over \Delta n}\right)^2
\ ,\label{e6}
\eeq
in the $(g_0,g_c)$ plane. In Eq.~(\ref{e6}),
$n_{SM}$ is the number of events expected for the \sm\
in one of the three channels (\ref{e3}--\ref{e5}),
$n(g_0,g_c)$ is the corresponding observed number of events
in the presence of anomalous couplings
and $\Delta n$
is the quadratic combination of statistical and systematic
error on the observed number of events
\beq
\Delta n = \sqrt{n(g_0,g_c) + (\epsilon n(g_0,g_c))^2}
\ .\label{e7}
\eeq
In the following, we take 1\%\ for the relative systematic error
$\epsilon$. This assumption is on the conservative side
considering the low \bkg\ environement of \ee\ collisions.
To evaluate the number of events
%$n={\cal L}\eta\sigma$
we assume an integrated luminosity of 10 fb$^{-1}$
and the following reconstruction efficiencies
for the reactions (\ref{e3}--\ref{e5}) respectively:
\beqa
\eta_{WW} & = & (1 - B(W\to\ell\bar\nu_\ell))^2
\approx .444 \ , \label{e8} \\
\eta_{ZW} & = & (1 - B(Z\to\nu\bar\nu) + B(Z\to\tau\bar\tau))
	(1 - B(W\to\ell\bar\nu_\ell))
\approx .513 \ , \label{e9} \\
\eta_{ZZ} & = & (1 - B(Z\to\nu\bar\nu) + B(Z\to\tau\bar\tau))^2
\approx .593 \ , \label{e10}
\eeqa
($B$ denotes \br s)
\ie,
we require both gauge bosons to decay with no missing energy\footnote{
	Because at lower orders no hadrons are produced in \ee\ collisions
	the ``gold-platted'' $W$ signature
	is not the leptonic
	but the hadronic one.
	This has to be contrasted with \ep\ or $p\bar p$ scattering.
}.
This way all the events can be fully reconstructed
and correctly identified as
belonging to one of the three reactions (\ref{e3}--\ref{e5}),
while no confusion is possible with lower order processes
like $\ee\to e^-\nu_eW^-$ or $\ee\to\ee Z^0$.
Since the gauge bosons tend to be emitted
rather isotropically
and with small energy,
their decay products have a very low probability
to escape detection through the beam pipe.
In contrast,
the bulk of the events in processes (\ref{e4},\ref{e5})
have their primary electrons\footnote{
	\ie\ the final state electrons
	which do not originate from the decay of a vector boson.}
deflected very little from the beam pipe.
Therefore,
to ensure that these primary electrons
are indeed seen under the experimental circumstances,
we impose the polar angle acceptance cut
\beq
10^o \le \theta_e \le 170^o
\ . \label{e11}
\eeq

Note that in reaction (\ref{e5})
only the combination $g_0+g_c$ of the couplings can be probed.
Therefore the contours reduce to two straight lines and
the 95\%\ confidence level is obtained
by setting $\chi^2 = 3.84$ in Eq.~(\ref{e6}) which corresponds
to a one-parameter fit.
In contrast,
the two parameter fits for the processes (\ref{e3},\ref{e4})
require $\chi^2 = 6$
to reach the same confidence level.

For each of the reactions (\ref{e4},\ref{e5})
we have only displayed
in Figs~\ref{f2}\ and \ref{f3}\
the combination of polarizations of the initial beams
which give the best limits.
The alternative combinations yield wider contours.
It appears
that the reaction (\ref{e4})
performed with two left-handed beams
yields the tightest bounds on the \aqgc.
Doubling the \cm\ energy
from 500 GeV to 1 TeV
improves the resolving power of the experiment
by almost a factor ten.
At the latter energy,
where several thousand events are expected \cite{ck},
even further improvements can be obtained
from processing also the information
contained in the angular distributions.

The region in the $(g_0,g_c)$ plane which can be explored
under similar conditions
in the \ep\ mode of a 500 GeV collider \cite{bb}
is also shown in Fig.~\ref{f2}.
Clearly,
stronger limits can be imposed on \aqgc\
when the information gathered from \ep\ and \ee\ scattering
is combined.

\vspace{-5mm}
\begin{ack}
\vspace{-5mm}
It is a pleasure to thank Edward Boos and Misha Dubinin
for a number of useful discussions
which triggered this work.
\end{ack}

\clearpage

\begin{figure}
\begin{center}
%%%%%%%%
% WWWW %
%%%%%%%%
\begin{picture}(100,100)
\Photon(20,20)(80,80){2}{10}
\Photon(20,80)(80,20){2}{10}
\Text(10,90)[c]{$W^+,1$}
\Text(90,90)[c]{$W^+,2$}
\Text(10,10)[c]{$W^-,3$}
\Text(90,10)[c]{$W^-,4$}
\end{picture}

$
g^2
\left[
	  2(1+g_c) \eta_{\mu_1\mu_2} \eta_{\mu_3\mu_4}
	- (1-g_c-2g_0)	\left(
				  \eta_{\mu_1\mu_3} \eta_{\mu_2\mu_4}
				+ \eta_{\mu_1\mu_4} \eta_{\mu_2\mu_3}
			\right)
\right]
$

\bigskip\bigskip\bigskip

%%%%%%%%
% WWZZ %
%%%%%%%%
\begin{picture}(100,100)
\Photon(20,20)(80,80){2}{10}
\Photon(20,80)(80,20){2}{10}
\Text(10,90)[c]{$W^+,1$}
\Text(90,90)[c]{$W^-,2$}
\Text(10,10)[c]{$Z^0,3$}
\Text(90,10)[c]{$Z^0,4$}
\end{picture}

$
-g^2 \cwt
\left[
	2 (1-\displaystyle{g_0\over\cwf}) \eta_{\mu_1\mu_2} \eta_{\mu_3\mu_4}
	- (1+\displaystyle{g_c\over\cwf})	\left(
				  \eta_{\mu_1\mu_3} \eta_{\mu_2\mu_4}
				+ \eta_{\mu_1\mu_4} \eta_{\mu_2\mu_3}
			\right)
\right]
$

\bigskip\bigskip\bigskip

%%%%%%%%
% ZZZZ %
%%%%%%%%
\begin{picture}(100,100)
\Photon(20,20)(80,80){2}{10}
\Photon(20,80)(80,20){2}{10}
\Text(10,90)[c]{$Z^0,1$}
\Text(90,90)[c]{$Z^0,2$}
\Text(10,10)[c]{$Z^0,3$}
\Text(90,10)[c]{$Z^0,4$}
\end{picture}

$
g^2
\displaystyle {g_0+g_c\over\cwf}
\left(
	  \eta_{\mu_1\mu_2} \eta_{\mu_3\mu_4}
	+ \eta_{\mu_1\mu_3} \eta_{\mu_2\mu_4}
	+ \eta_{\mu_1\mu_4} \eta_{\mu_2\mu_3}
\right)
$
\end{center}
\bigskip
\caption{Feynman rules for the three quartic gauge interactions
	which are modified or generated
	by the addition of the anomalous pieces
	(\protect\ref{e1},\protect\ref{e2})
	to the \sm\ lagrangian.}
\label{f1}
\end{figure}

\clearpage

\begin{figure}[htb]
\centerline{\input /home/iws166/cuypers/anom/500.tex}
\bigskip
\caption{Contours of observability at 95\%\ confidence level
	of the \aqgc\ $g_0$ and $g_c$.
	The experiments (\protect\ref{e3}--\protect\ref{e5})
	are performed at 500 GeV
	with 10 fb$^{-1}$ of accumulated luminosity.
	The limits which can be obtained
	under similar conditions
	in the \ep\ mode of the same collider
	\protect\cite{bb}
	are indicated by the thin line.}
\label{f2}
\end{figure}

\clearpage

\begin{figure}[htb]
\centerline{\input /home/iws166/cuypers/anom/1000.tex}
\bigskip
\caption{Same as Fig.~\protect\ref{f2}
	at 1 TeV.}
\label{f3}
\end{figure}

\end{document}